\documentclass[twocolumn]{aastex63}

\usepackage{wrapfig}
\usepackage{tcolorbox}
\usepackage{stix}
\usepackage{appendix}
\usepackage[utf8]{inputenc}
\usepackage{floatflt}

\begin{document}

\title{Probing the Taxonomic Diversity of the Small Near-Earth Object Population}

\correspondingauthor{Andy J. L\'{o}pez-Oquendo}
\email{andy.lopez-oquendo@nasa.gov}

\author[0000-0002-2601-6954]{Andy J. L\'{o}pez-Oquendo}
\affiliation{NASA Goddard Space Flight Center, Solar System Exploration Division, Greenbelt, MD 20771, USA}
\affiliation{Department of Astronomy and Planetary Science, Northern Arizona University, Flagstaff, AZ 86011, USA}

\author[0000-0003-4580-3790]{David E. Trilling}
\affiliation{Department of Astronomy and Planetary Science, Northern Arizona University, Flagstaff, AZ 86011, USA}

\author[0000-0002-8132-778X]{Michael Mommert}
\affiliation{Stuttgart Technical University of Applied Science, Stuttgart, Germany}

\author[0000-0001-8617-2425]{Micha\"{e}l Marsset}
\affiliation{European Southern Observatory199, Karl-Schwarzschild-Strasse 2, 85748 Garching bei München, Germany}

\begin{abstract}

Small near-Earth objects (NEOs) represent the immediate source population of meteors and meteorites and provide a direct window into the compositional and dynamical evolution of the inner Solar System. However, their small sizes, rapid rotation rates, and short observability windows have historically limited systematic compositional characterization. We present results from a multi-year rapid-response near-infrared spectrophotometric survey of recently discovered NEOs conducted with the UKIRT–WFCAM instrument between 2016 and 2024. Our campaign obtained 371 observation sequences, yielding reliable color measurements and probabilistic taxonomic classifications for 122 NEOs, with diameters ranging from $\sim$5~m to $\sim$150~m and a median size of $\sim$60~m. We employed lightcurve-corrected $Z$, $J$, $H$, and $K$ photometry and a machine-learning–based classification framework to derive both observed and debiased taxonomic distributions as a function of object size. The observed taxonomic fractions shows a population of small NEOs dominated by S-complex and C/X-complex objects at almost equal proportions. After accounting for observational biases, however, the inferred intrinsic population is instead dominated by dark, carbonaceous (C/X-complex) bodies, with implications for the true compositional makeup of the small NEO population. We further explore connections between small NEOs, meteorite fall statistics, and atmospheric filtering effects, finding that a substantial fraction of carbonaceous material likely fails to survive atmospheric entry. These results highlight the importance of debiased surveys of small NEOs for interpreting the meteorite record, refining delivery models from the main belt, and informing future mission target selection and planetary defense strategies.

\end{abstract}

\section{Introduction \label{sec:intro}}

Small near-Earth objects (NEOs) are the likely progenitors of meteors and meteorites, whose collection has provided a wealth of chemical information about the Solar System and objects that are otherwise inaccessible in space. Thanks to extensive spectroscopic efforts, it is now well established that ordinary chondrites are compositionally linked to S-complex asteroids \citep{GAFFEY1993573, THOMAS2010419, Nakamura2011, Tsuchiyama2011, Binzel2019}, carbonaceous chondrites to C-complex asteroids \citep{BOWELL1978313, VILAS1985503, Vernazza_2015}, aubrites or enstatite-rich meteorites to X-complex asteroids \citep{GAFFEY199295, Fornasier2008, Fornasier2011, Keil2010}, and Howardite–Eucrite–Diogenite (HED) meteorites to the asteroid Vesta \citep{CRUIKSHANK19911, Binzel1993, Russell2012}. Despite these well-established links, much remains to be explored within the small, observationally challenging population of NEOs.

Sub-kilometer NEOs differ fundamentally from their larger counterparts in that they are far more numerous, collisional-evolved, more sensitive to non-gravitational forces, and often exhibit rapid rotation and irregular shapes, factors that can strongly influence both their physical properties and observational detectability \citep{PRAVEC200012, Delbo2007, POLISHOOK20149}. These characteristics, combined with their intrinsic faintness and short discovery-to-observation timescales, make small NEOs particularly challenging to characterize spectroscopically, requiring rapid-response observing strategies to mitigate lightcurve and selection effects \citep{Mommert2016, Erasmus2017, Lopez-Oquendo2023}. As a result, existing taxonomic surveys are inherently biased toward brighter, higher-albedo, and more slowly rotating objects, potentially skewing the inferred compositional distribution of the near-Earth population \citep{Harris2015, Heinze_2021, Broz2024}. With the advent of next-generation discovery facilities such as the Vera C. Rubin Observatory and NEO Surveyor, the number of detected small NEOs will increase dramatically, further emphasizing the need for robust, bias-aware taxonomic characterization frameworks \citep{Petit2011, Marsset_2022}.

Establishing robust compositional and dynamical connections between asteroids and meteorites, and quantifying their relative abundances in near-Earth space, remains a central topic in planetary science. Understanding which objects are most common in Earth’s vicinity has important implications for planetary defense \citep{Reddy_2024}, delivery mechanisms from the main belt \citep{Vokrouhlick2000, MORBIDELLI2003120, Bottke2006,Broz2024,Brovz2024b}, mission planning and target selection \citep{Lauretta2023}, interpretation of the meteorite record \citep{Greenwood2020, Gattacceca2025}, and studies of evolutionary processes \citep{Delbo2007, Shober2025}. Meteorite fall statistics indicate that approximately $\sim$80\% of falls originate from ordinary chondrites, followed by $\sim$8.2\% achondrites (including HEDs), $\sim$4.4\% carbonaceous chondrites, $\sim$1.6\% enstatite chondrites, and $\sim$5.6\% iron and stony-iron meteorites \citep{Harvey1989, Gattacceca2025}. However, such statistics cannot be directly extrapolated to the intrinsic population of small NEOs, as atmospheric entry significantly modifies the surviving fall distribution at an uncertain rate \citep{Shober2025}.

From an observational perspective, spectroscopic \citep{Devogele2019, Binzel2019, Marsset_2022, Sanchez2024} and spectrophotometric surveys \citep{Mommert2016, Erasmus2017, Navarro2024, Moskovitz_2026} have attempted to address which taxonomic classes dominate the NEO population. These large-scale visible and near-infrared surveys, while not yet conclusively robust, consistently suggest that silicate-rich (S-complex) objects dominate near-Earth space. Such findings raise the fundamental question of why the most abundant objects in the main belt are not similarly represented among NEOs. Although classical dynamical models of asteroid delivery \citep{MORBIDELLI2003120, Bottke2006,Granvik2018,Nesvorny2023,Broz2024} demonstrate that transport efficiencies depend strongly on source-region location—particularly proximity to powerful resonances—obtaining an intrinsic census of the small NEO population is essential for refining these models and improving our understanding of NEO evolutionary pathways.

In light of the importance of accessing and characterizing the smaller members of the NEO population (i.e., diameters $<300$~m), and the need for a more comprehensive quantitative survey of these objects, we present a multi-year rapid-response taxonomic campaign. This work extends the early efforts of \citet{Mommert2016} by including a substantially larger sample of smaller objects, improved photometric acquisition strategies, and a bias-corrected analysis. The thousands of hours of observations described in Section~\ref{sec:meto_data_analysis} enabled the collection of spectrophotometric data for 230 NEOs. In this study, we investigate both the observed and debiased taxonomic distributions of small NEOs and their dependence on object size. We also place constraints on NEO rotational properties and limits on lightcurve amplitudes (Section~\ref{sec:results}). In Section~\ref{sec:debias}, we describe the debiasing approach and discuss the dominant bias factors affecting this and similar taxonomic surveys. Finally, we examine the implications for NEO source regions, meteorite connections, and atmospheric filtering processes (Sections~\ref{sec:results} and \ref{sec:discussion}) in the context of both observed and debiased taxonomic fractions.

\section{Data Acquisition and Analysis \label{sec:meto_data_analysis}}

\subsection{Target Selection and Observations \label{sec:observation}}
In this program, we acquired rapid-response near-infrared spectrophotometry of recently discovered NEOs between 2016 and 2024 using the Wide-Field Camera \citep[WFCAM;][]{Casali2007} mounted on the 3.8~m United Kingdom Infrared Facility (UKIRT, now also known as the University of Hawai`i Infrared Telescope)\footnote{\url{https://about.ifa.hawaii.edu/ukirt/}} located at Maunakea, Hawai`i. Recently discovered NEO orbits are typically confirmed within days of their close approach to Earth, or near the time when they reach peak brightness. After their encounter with Earth, these objects rapidly fade in brightness \citep{GALACHE2015}. Consequently, such NEOs require fast-response observing strategies in order to be observed near or prior to their closest approaches. The UKIRT telescope is particularly well suited for short-notice or rapid-response observations, as it operates in queue mode. In this mode, telescope operators execute approved and priority-ranked Minimum Schedulable Blocks (MSBs) submitted through the UKIRT Observation Management Project (OMP) portal\footnote{\url{https://about.ifa.hawaii.edu/ukirt/omp-web-page-contents-and-usage/}} using the Observing Tool (OT).

To submit MSBs, we used a Python script to query the Minor Planet Center (MPC)\footnote{\url{https://minorplanetcenter.net/data}} on a daily basis. First, the script downloads the most up-to-date MPC NEO list, which contains all known near-Earth asteroids. The script then performs a cleaning process and retrieves newly discovered NEOs identified within the five weeks preceding the execution date. We discard objects with visible magnitudes {$V \geq 22.0$~mag} and with Galactic latitude $|b| < 15^{\circ}$ because, as explained in our previous work \citep{Mommert2016}, $Z$-band calibration from SDSS $z$ is unreliable in these regions. For objects meeting our selection criteria, we compute their observability using the JPL Horizons system\footnote{\url{https://ssd.jpl.nasa.gov/horizons/}} \citep{Giorgini_2015}, requiring an airmass $\leq 2.0$. Once the final target list is assembled, we sort the NEOs by visible magnitude and prioritize those with larger $H_{V}-V$ values to ensure observations are obtained {their peak in brightness}. The script then generates an MSB containing the NEOs’ orbital parameters from JPL Horizons along with additional observing information (e.g., integration times, number of frames, and filter sequences), which is subsequently uploaded manually to the UKIRT OMP.

Observations were conducted by acquiring $Z$, $J$, $H$, and $K$ photometry to determine the $Z-J$, $J-H$, and $J-K$ colors of the NEOs. Because the primary goal of this program was to characterize very small NEOs, we adopted an observing strategy designed to minimize variability in taxonomic classification arising from target shape irregularities. Small NEOs are known to exhibit highly non-spherical shapes due to their rubble-pile nature, and as a result, color measurements can be affected by rotational lightcurve variations. Obtaining lightcurve-corrected colors is therefore critical for reliable taxonomic classification using spectrophotometry. To mitigate the effects of rotational brightness variability, we interspersed each non-$J$-band exposure with a $J$-band observation, as objects are typically brightest in this band, yielding higher signal-to-noise ratios. For example, a typical filter sequence for objects brighter than $V \leq 19.5$~mag consisted of JKJZJHJKJ. For fainter objects ($V \geq 20.5$~mag), we employed a longer sequence of JKJZJKJHJZJKJHJKJ. In some cases, though infrequently, objects were observed in multiple consecutive cycles. We attribute these instances to the absence of MSBs from other projects, which resulted in additional observing time being available during the night.

\begin{figure}[!] 
    \centering 
    \includegraphics[width=0.47\textwidth]{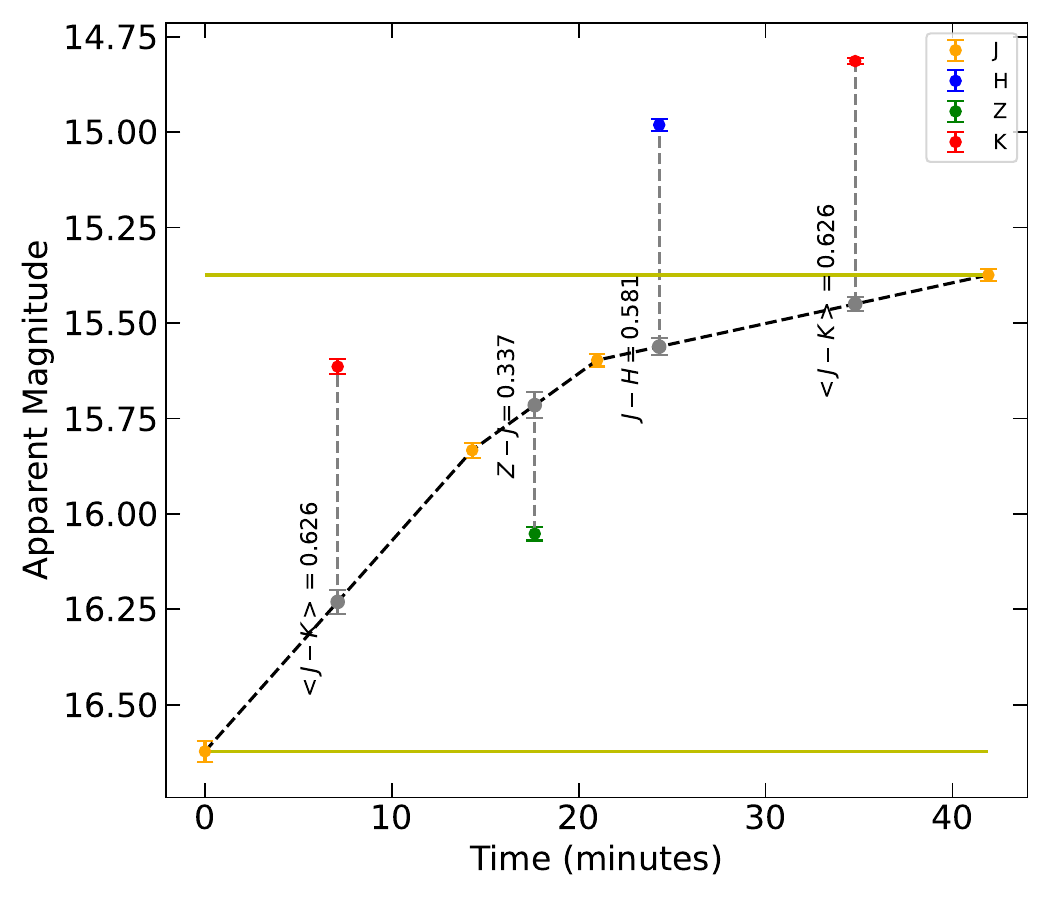} \caption{Example of our lightcurve correction approach and color determination for NEO 2022~VV2. The dashed black line illustrates the interpolated J-band lightcurve and the overplotted grey circles show the predicted magnitude, which is used to compute the lightcurve corrected color at the time of a non-J-band observation. {Each colored circle represents the binned magnitude obtained by combining all photometric measurements acquired in the corresponding filter.} The horizontal yellow lines indicate the lightcurve amplitude of the target across the observing sequence. The colors shown in the figure have not been solar color-subtracted. That is, each respective color contains the $Z-J=0.369$, $J-K=0.304$, and $J-H=0.354$~mag color of the Sun. The variability timescale is measured from the time difference at which the lightcurve reaches the respective minima and maxima during the range of observation. In this case, although a full lightcurve was not obtained, a confident argument that the object has a rotation period $>$40~minutes can be made.} \label{fig:lc_color} 
\end{figure}

We acquired UKIRT photometry using non-sidereal tracking as described in our initial approach \citep{Mommert2016}. Individual exposure times were fixed at 5~s in all filters to minimize trailing losses while tracking the asteroid, as most targets moved at rates of $\lesssim5$ arcsec~min$^{-1}$. Rather than varying the exposure time on a target-by-target basis, we adjusted the total integration time by modifying the number of acquired frames according to the target brightness. The required integration time was estimated using the UKIRT exposure time calculator and chosen to achieve a signal-to-noise ratio of at least $\sim10$ in each filter.

In contrast to \citet{Mommert2016}, we increased the number of frames acquired in the $Z$, $H$, and $K$ bands to improve the reliability of the measurements. In general, an observing block lasted between 40~minutes and 2~hours, yielding a total of 32, 80, 32, and 196 frames in the $Z$, $J$, $H$, and $K$ bands, respectively, for objects with $V{\leq}19.5$. For objects with $V{\geq}20.5$, the number of collected frames in each band increased by approximately a factor of three. We obtained substantially more frames in the $K$ band to compensate for its poorer performance, which results from higher sky background, reduced detector sensitivity, and lower solar flux between 2.0 and 2.4~$\mu$m. {In Figure \ref{fig:lc_color} we show an example of the binned photometry at each band for NEO 2022~VV2.} Unless noted otherwise, additional observing details are provided in \citet{Mommert2016}.

\subsection{Data reduction and Color Measurements \label{sec:reduction}}

The UKIRT WFCAM data are reduced by the Cambridge Astronomical Survey Unit (CASU) using the procedures described by \citet{Irwin2007} and are subsequently ingested into the UKIRT Science Archive. The reduced data are downloaded from the WFCAM archive, and each observed field, corresponding to the filter sequences detailed in \S\ref{sec:observation}, is split into individual datasets for each target. We then perform aperture photometry using the Python-based Photometry Pipeline software \citep[PP;][]{MOMMERT201747}. PP performs image registration using the 2MASS star catalog to identify sources in the field and combines the stationary sky frame (\textit{skycoadd}) and the moving asteroid frame (\textit{comove}) using Source Extractor (SEx), SCAMP, and SWARP \citep{Bertin1996, Bertin2002, Bertin}. PP calls SEx to perform aperture photometry on each image within a field. The combination of the \textit{comove} and \textit{skycoadd} frames helps to avoid potential stellar contamination in the final asteroid photometry.

\begin{figure}[!]
    \centering
    \includegraphics[width=0.47\textwidth]{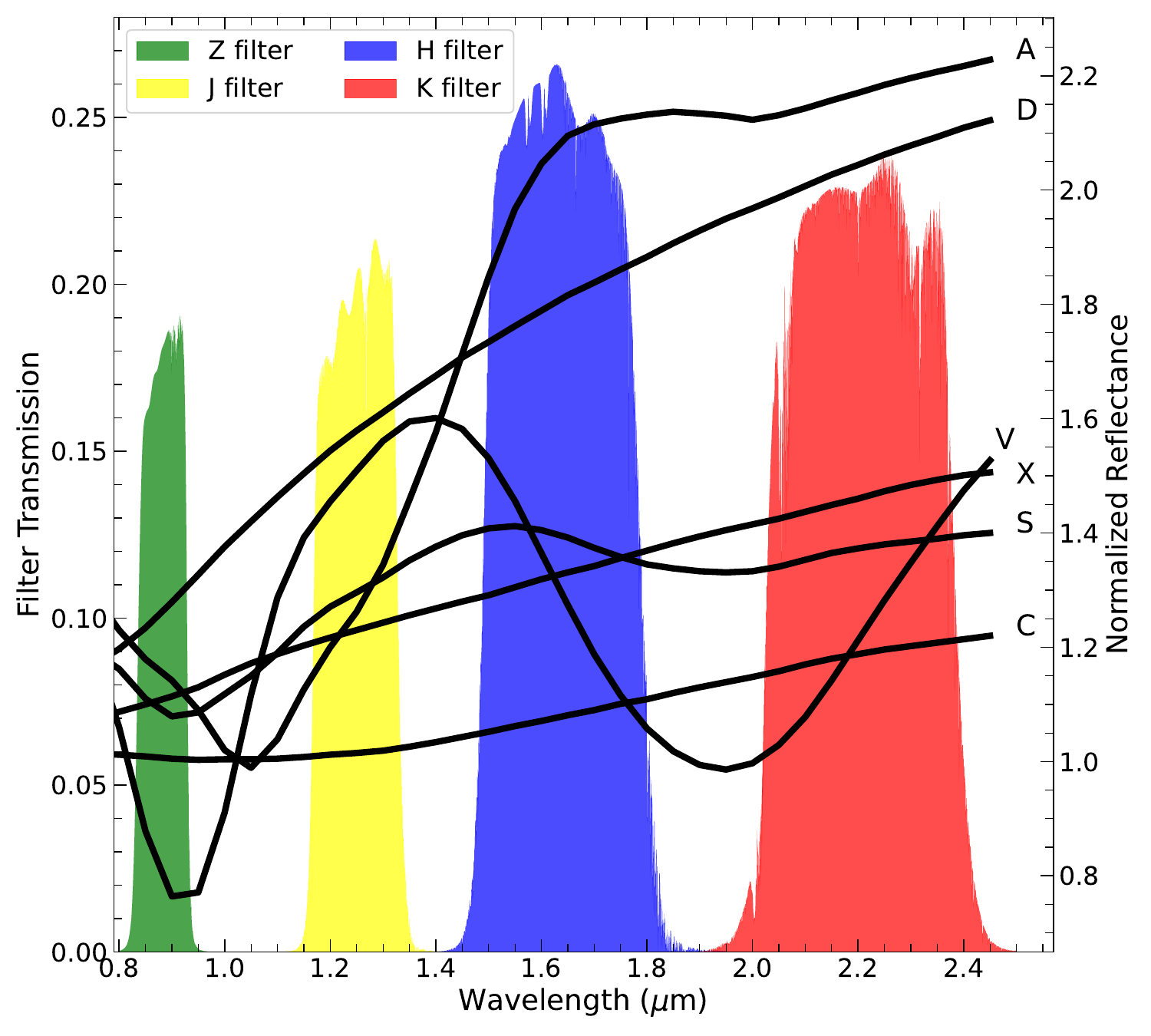}
    \caption{UKIRT WFCAM response functions for the $Z, J, H,$ and $K$ filters as a function of wavelengths. The filter transmission values are displayed on the left y-axis. The black lines show the average reflectance at (right y-axis) for the Bus-DeMeo taxonomic classes used in our classification scheme. The reflectance was normalized at 0.55~$\mu$.}
    \label{fig:filters}
\end{figure}

To obtain the asteroid magnitudes in each band, we binned the collected magnitudes within each bandpass and computed weighted photometric uncertainties. To derive lightcurve-corrected colors, we first interpolated the $J$-band lightcurve and obtained the corresponding magnitude at the time of each non-$J$-band binned measurement. We then computed the differences in band magnitudes to derive the $J-K$, $Z-J$, and $J-H$ colors. The color uncertainties were estimated by propagating the uncertainty of the non-$J$-band magnitude with the average uncertainty of the $J$-band measurements immediately before and after the non-$J$-band observation. In Figure~\ref{fig:lc_color}, we show an example of a lightcurve-corrected color measurement for NEO 2022~VV2, highlighting the importance of correcting for rotational brightness variability, particularly for highly elongated NEOs. For asteroids with multiple measurements of a given color, particularly $J-K$, we adopted the average value (i.e., $\langle J-K\rangle$, $\langle J-H\rangle$, and $\langle Z-J\rangle$).

We measured the lightcurve amplitude ($A^{*}$), which we defined as the difference between the $J$-band lightcurve minima and maxima. We also computed the variability timescale ($\tau^{*}$), which we defined as the difference between the time at $J$-band lightcurve minima and maxima. As shown in Figure~\ref{fig:lc_color}, in many cases only a portion of the lightcurve is sampled; therefore, these estimates should be treated as lower limits on the object’s elongation and rotation period. The amplitude uncertainty was estimated by propagating the uncertainties of the $J$-band photometric minima and maxima. Figure~\ref{fig:lc_color} illustrates an example of this approach.

\subsection{Probabilistic Taxonomy Classification}
\subsubsection{Construction of Training Sample \label{sec:training}}

Asteroid spectral properties, such as albedo, spectral slope, and absorption features, are captured in broadband photometry and are directly reflected in an asteroid’s colors. The $Z$, $J$, $H$, and $K$ broadband filters cover wavelength ranges in which characteristic compositional spectral features of asteroids occur. In Figure~\ref{fig:filters}, we present an example showing the average spectra of different asteroid taxonomies from the \citet{Demeo2009} scheme together with the corresponding WFCAM NIR filter response functions. Although spectrophotometry does not provide the same level of spectral detail as spectroscopy, discrimination between the main complexes and distinct spectral types (e.g., D-type, S-complex, C/X-complex, and V-type) is possible. In some cases, however, this approach may yield false-positive classifications, depending on specific regolith spectral properties that produce similar integrated reflectance within a given bandpass. For example, a positively sloped S-type spectrum and a flatter, featureless spectrum that compensates for olivine and pyroxene absorption features can result in similar broadband fluxes.

To taxonomically classify the measured asteroid colors, we used the MIT–Hawai`i Near-Earth Objects Spectroscopic Survey (MITHNEOS), the largest available spectral database of NEOs \citep{Binzel2019}, as ground-truth guidance. We retrieved 1421 asteroid spectra (including repeated observations) from the MITHNEOS database and classified them using the \citet{Demeo2009} taxonomy classification tool\footnote{\url{http://smass.mit.edu/busdemeoclass.html}}. We selected spectra with an absolute average residual $\leq$0.05 and residuals within 10\% of the minimum residual value. This strict selection yielded a robust sample of 689 spectra, which helped reduce redundant classifications and improved the efficiency of the synthesized training sample.

\begin{figure}[!]
    \centering
    \includegraphics[width=0.47\textwidth]{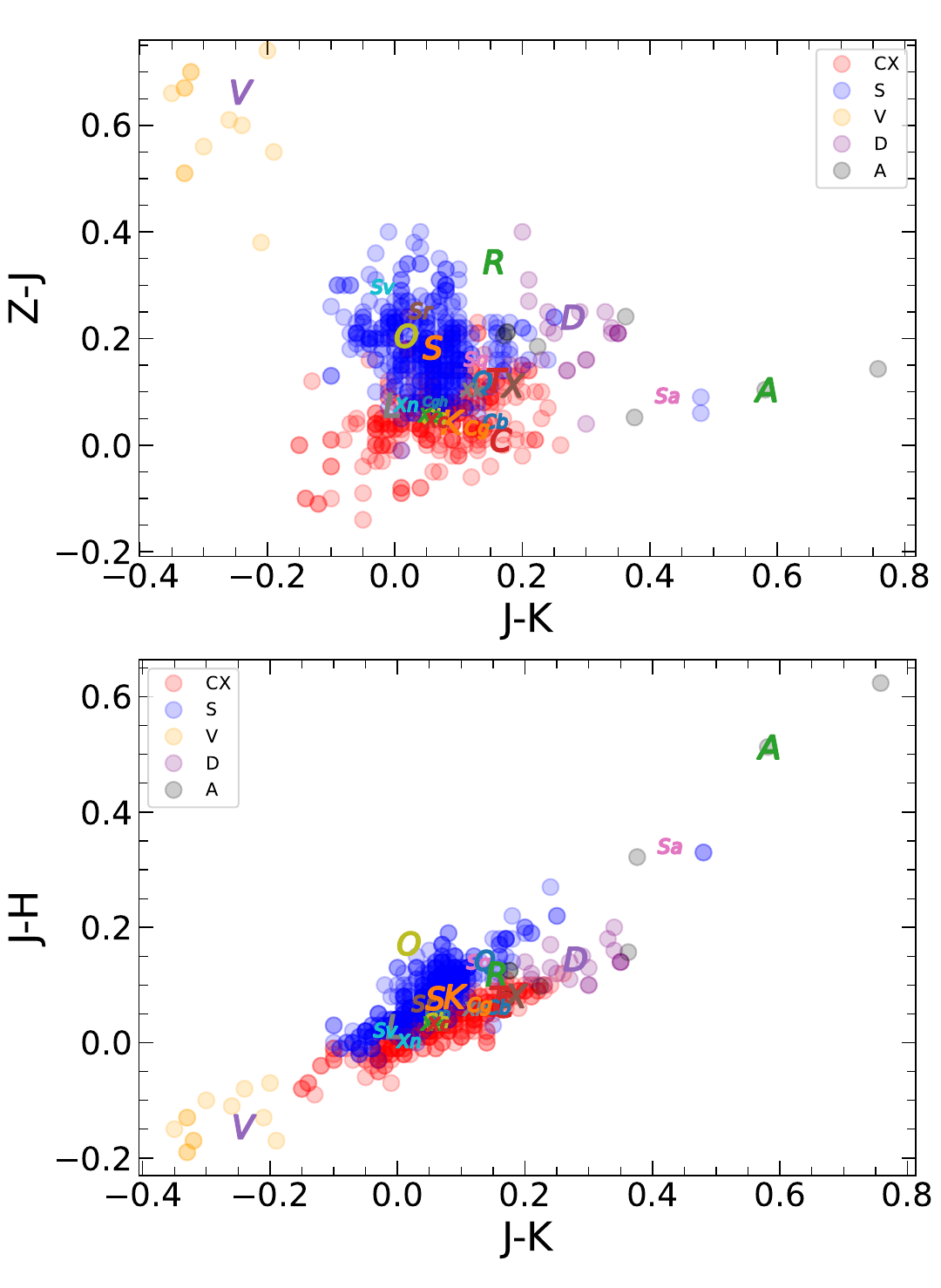}
    \caption{Color-color scheme of our training sample using ground-truth spectra from the MITHNEOS survey. The top panel shows $Z-J$ versus $J-K$ color of taxonomically classified C-/X-, S-, D-, V-, and A-types asteroids in red, blue, purple, yellow, and gray circles. The bottom panel shows the $J-H$ versus $J-K$ color for the before-mentioned taxonomies. These synthesized spectra colors have been solar-color removed and they reside in the Vega-magnitude system. The over-plotted letters show the colors of the respective average spectra from \citet{Demeo2009}. Note that in our color-color system, the distinction between C- and X-complexes is hardly noticeable. Thus, we clustered the C and X-complex into a similar C/X space to minimize redundant classifications.}
    \label{fig:training}
\end{figure}

Using the UKIRT filter response functions shown in Figure~\ref{fig:filters}, we derived the corresponding $J-K$, $Z-J$, and $J-H$ colors for each classified spectrum using a spectral convolution approach \citep{Lopez-Oquendo2022, Lopez-Oquendo2024}. Figure~\ref{fig:training} shows the resulting color–color training sample used to classify the measured UKIRT-WFCAM NIR colors. The colors displayed in Figure~\ref{fig:training} are solar-subtracted and expressed in the Vega magnitude system. The training sample includes 400 S-complex, 259 C-/X-complex, and 30 V- plus D-types synthesized asteroid colors. As expected, objects with strong absorption features (i.e., V-types) are clearly distinguished from other taxonomic classes. In color–color space, featureless spectra such as those of the X- and C-complexes are difficult to distinguish; therefore, we group them together, as the precision of our photometry does not allow reliable separation. Another notable feature of the color–color ground-truth distribution is that S- and C-/X-complexes partially overlap in $J-K$, $Z-J$, and $J-H$, reflecting the effects of space weathering and mineralogical diversity (e.g., variations in albedo, absorption bands, and spectral slopes) within these taxonomies. Consequently, objects located near the boundaries of these color distributions may yield ambiguous or multiple classifications (see Section~\ref{sec:classifying}).

\subsubsection{Classifying Colors} \label{sec:classifying}

We evaluate the performance of the classifier (i.e., Gaussian Naive Bayes and k-nearest neighbor) using the accuracy, precision, recall, and F1-score metrics for each experimental configuration. We evaluate our trained k-Nearest Neighbor (k-NN) model using a range of metrics. While accuracy depicts the overall correctness of all model predictions, precision measures the fraction of predicted objects belonging to a given class that are correctly classified and recall addresses the completeness of our predictions; F1-score represents the harmonic mean of precision and recall, providing a simplified means to assess the overall model performance. Metrics quoted in Table~\ref{tab:accuracy} are based on a leave-one-out evaluation strategy for different values of the hyperparameter k, the number of nearest neighbors considered in the prediction. We find model predictions based on three colors to perform highly under all metrics; k only has a minor impact on the results. In the case of two colors, our k=3 model slightly outperforms the k=5 model, but overall performance is clearly lower than in the three-color case. A significant drop in performance occurs in the single-color case. This trend demonstrates that the inclusion of multiple color indices provides important discriminatory power for the classification task. However, in all cases, precision and recall show a somewhat balanced behavior, indicating that dataset imbalance does not pose a significant issue to our classification problem. We deliberately reject classifications with k=1, which makes the model more susceptible to overfitting. Therefore, we limit our analysis to the case k$\geq$3. Using k=1 would lead to a higher fraction of S-complex and fewer C-/X-complex and D-types, which may well be an overfitting effect.

\begin{deluxetable}{ccccccc}
\tablecaption{K-nearest neighbor accuracy, precision, recall, and F1 scores for three training configurations using three, two, and one color indices.\label{tab:accuracy}}
\tablehead{Colors & Method & k & Accuracy & Precision & Recall & F1}
\startdata
Three & k-NN & 3 & 0.95 & 0.98 & 0.98 & 0.98 \\
Three & k-NN & 5 & 0.96 & 0.94 & 0.94 & 0.94 \\
Two & k-NN & 3 & 0.88 & 0.87 & 0.81 & 0.84 \\
Two & k-NN & 5 & 0.86 & 0.85 & 0.8 & 0.82 \\
One & k-NN & 3 & 0.65 & 0.57 & 0.46 & 0.5 \\
One & k-NN & 5 & 0.72 & 0.57 & 0.46 & 0.5 \\
Three & GNB & - & 0.92 & 0.9 & 0.86 & 0.88 \\
Two & GNB & - & 0.87 & 0.92 & 0.85 & 0.88 \\
One & GNB & - & 0.69 & 0.45 & 0.46 & 0.45 \\
\enddata
\end{deluxetable}

To provide a probabilistic taxonomic classification, we applied the selected algorithms to individual NEOs and explored a range of possible classifications within the photometric color uncertainties. We assumed a Gaussian uncertainty distribution centered on each measured color with a 1-$\sigma$ standard deviation. For each object, we generated $10^{6}$ Monte Carlo realizations (“clones”) of the colors and evaluated the classification likelihood across this ensemble. The probabilistic taxonomy was then determined by measuring the relative frequency of each taxonomic class within the $10^{6}$ trials. Finally, we assigned the resulting percentages for the S-, C-/X-complex, V-, D-, and A-types classes to each object, as reported in Table~\ref{tab:results}.

\begin{figure*}[!]
\centering
\includegraphics[width=0.8\textwidth]{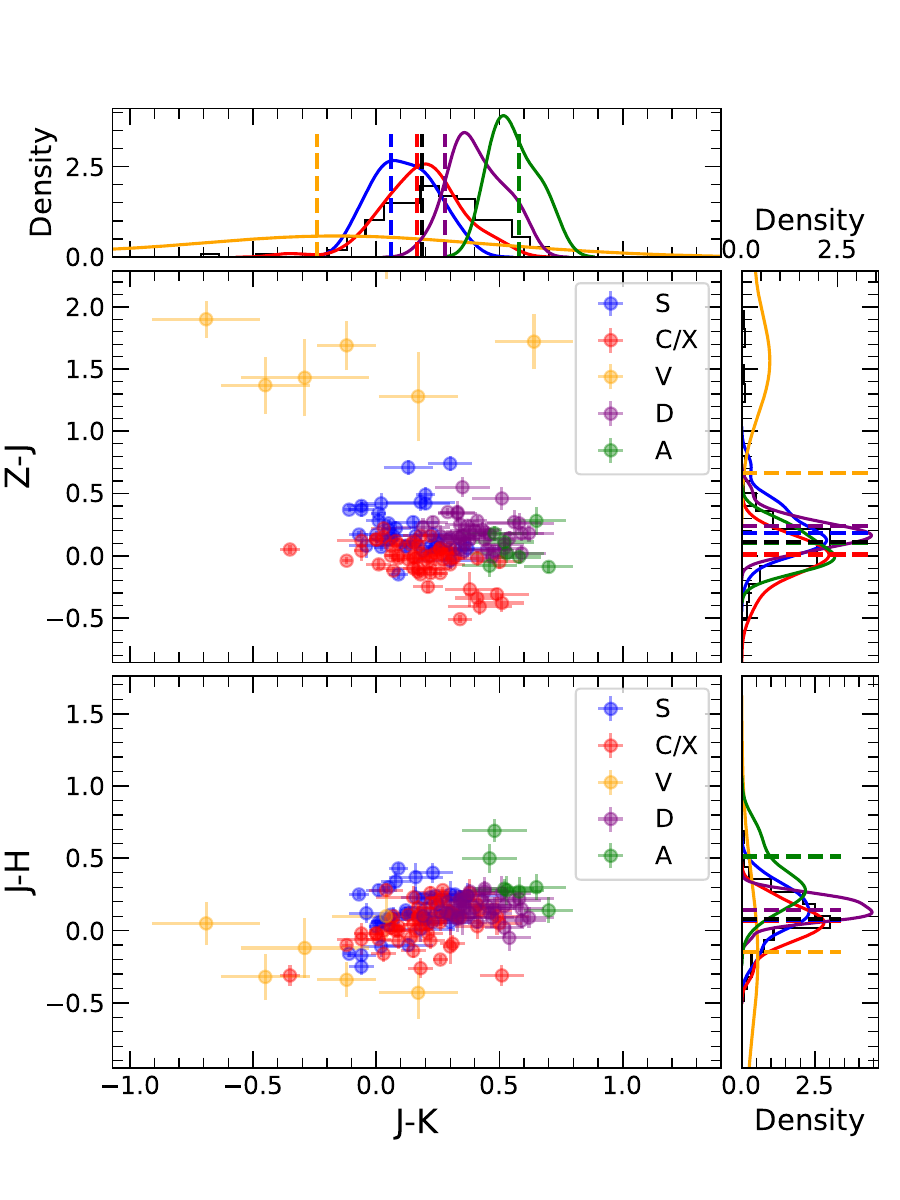}
\caption{Color–color diagram showing $Z-J$ (top) and $J-H$ (bottom) as a function of $J-K$ for all classified objects in the UKIRT-NEO survey. Colors and uncertainties are colored according to the object’s taxonomic classification. The top and rightmost subpanels show the kernel density distributions for each color for objects classified as C/X (red), S (blue), V (yellow), D (purple), and A (green). Vertical dashed lines indicate the colors of the average taxonomy spectra.}
\label{fig:colors}
\end{figure*}

\section{Results \label{sec:results}}

We performed 371 observations of recently discovered NEOs between 2016 and 2024. Of these, 28 observations could not be processed by the photometry pipeline due to poor data quality (i.e., low SNR, poor weather, or incomplete observations), leaving 343 observations for analysis. From this subset, we obtained probabilistic taxonomic classifications for 259 observations of 230 unique NEOs. Reliable classifications were achieved {for 122 unique NEOs, defined as those with a dominant taxonomic probability larger than 50\% and a total photometric uncertainty $\leq$0.2~mag}. The remaining objects resulted in a non-definite solution, given that the probabilities were nearly equal among different taxonomic classes due to the large photometric uncertainty. The mean lightcurve amplitude in our sample is 0.47~mag, with a mean variability timescale of approximately 28~minutes, suggesting a somewhat elongated shape for most of the surveyed small NEOs. The 28~minute variability timescale may reflect a sampling artifact; nevertheless, these estimates should be treated as lower limits. All the details about the probabilistic classification, amplitude, variability timescale and measured colors for individual objects (both robust and non-definite classifications) are listed in Table \ref{tab:results}.

Figure~\ref{fig:colors} presents the color–color distribution based on the measured $Z-J$, $J-K$, and $J-H$ colors of the classified objects in our sample (without subtraction of solar colors). A subset of targets exhibits inconsistencies among the measured colors, which may introduce biases in the resulting taxonomic classifications. Given the highly irregular shapes and rapid rotation rates expected for very small NEOs, our lightcurve-correction scheme may not fully account for complex variations in reflected flux with changing viewing geometry. The apparent bimodality observed in the $J-K$ color distribution could arise from these effects, in combination with additional photometric systematics discussed in Section~\ref{sec:survey_accuracy}.

\begin{figure*}[!]
\centering
\includegraphics[width=0.7\textwidth]{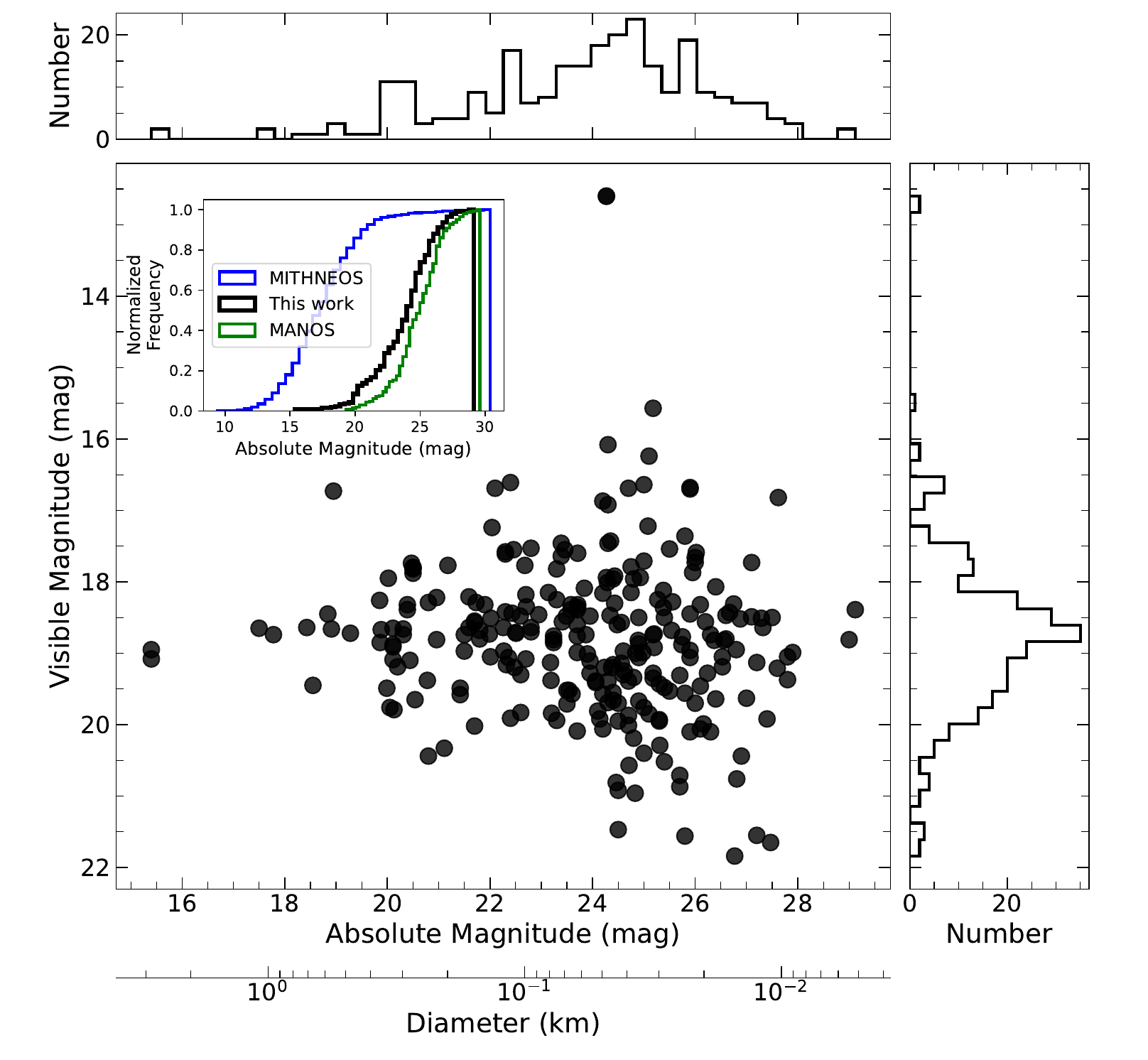}
\caption{Visible magnitude at the time of observation as a function of absolute magnitude $H_{V}$ for the UKIRT-NEOs surveys. The top axis shows the estimated diameter derived from $H_{V}$ assuming a 15\% albedo using the relation of \citet{Harris2002}. Most objects were observed at $V<H_{V}$, highlighting the efficiency of the UKIRT-NEO survey in targeting NEOs near peak in brightness. The bottom panel shows the cumulative $H_{V}$ distributions for the UKIRT-NEO survey (black line), MANOS (green line), and MITHNEOS (blue line). The dashed black line indicates the marginal $H_{V}$ distribution of the UKIRT-NEO survey.}
\label{fig:HV}
\end{figure*}

Our survey includes objects with diameters ranging from 2.8~km down to 5~m. As shown in Figure~\ref{fig:HV}, approximately 80\% of the targets have $H_{V}\geq22$~mag, corresponding to diameters smaller than 130~m, and about 50\% have diameters below 55~m. The median absolute magnitude of the survey is $H_{V}=24.24$~mag (approximately 55~m in diameter), with a mean value of $H_{V}=23.85$~mag (approximately 63~m in diameter). Figure~\ref{fig:HV} further shows that most targets were observed at apparent magnitudes $V \leq H_{V}$, demonstrating the effectiveness of this rapid-response campaign in observing recently discovered NEOs near or at their close approach.

\section{Discussion \label{sec:discussion}}

\subsection{Classification accuracy and limitations \label{sec:survey_accuracy}}

In Table~\ref{tab:accuracy} and Figure~\ref{fig:confusion_matric}, we demonstrate the robustness of our training sample to taxonomically classify an object with near-IR photometry. The confusion matrix shown in Figure~\ref{fig:confusion_matric} provides further insight into the classification performance and reveals systematic trends in the misclassifications of taxonomic classes. {In the three-color configuration, around 33\% of C-complex objects could falsely classify as D-types while about 50\% and 25\% of A-types could classify as either S and C/X, respectively. As the number of available colors decreases, the confusion between classes increases. In particular, objects belonging to intermediate or neutral classes become increasingly scattered toward the red end of the color space, resulting in a higher fraction of D-type false positives.}

{The confusion matrix therefore suggests that the dominant source of classification error is not purely random, but instead reflects the geometry of the color space combined with the photometric uncertainties of the survey. Importantly, this effect does not imply that the reddest objects in our sample are intrinsically D-types. Rather, it indicates that noisy measurements preferentially move objects into the region occupied by D-types in our adopted classification scheme, thereby increasing the probability of false-positive D classifications.}

{A plausible observational mechanism contributing to this asymmetry arises from the combination of the survey cadence, photometric uncertainties, and the rotational variability of the targets. As discussed in Section~\ref{sec:meto_data_analysis}, our observations consist of sequences of 5~s exposures acquired while UKIRT tracks the asteroid. In principle, photometric uncertainties alone would be expected to scatter the measured colors in both directions. However, for rapidly rotating and highly elongated objects, frames obtained near the fainter portions of the rotational lightcurve generally have lower signal-to-noise ratios and may therefore be rejected by the photometry pipeline or assigned lower statistical weight in the final averaged magnitude. Consequently, the measured magnitude can become preferentially weighted toward observations obtained when the object is brighter. This effect is expected to be most significant in the $K$ band, where the survey performance is poorer because of the higher sky background and lower detector sensitivity. As a result, the measured $K$-band magnitude may become systematically brighter, producing artificially redder $J-K$ colors. Such a bias would increase the likelihood of assigning D- or A-type classifications, even when the intrinsic colors correspond to another taxonomic class. This effect is expected to be most relevant for the faintest targets in our survey, for which the $K$-band measurements were closest to the detection limit.}

{We investigated the reproducibility of our taxonomic classifications by analyzing objects with multiple sets of sparse observations (i.e., observations obtained on different days) and probabilistic taxonomies greater than 0.5, resulting in a sample of sixteen candidates. Eight objects (2019~SC9, 2019~VH5, 2019~WB7, 2020~BE2, 2020~RF, 2020~UN3, 2021~DX1, and 2022~UN) yielded consistent probabilistic classifications across all observing epochs. In particular, NEOs 2021~DX1 (S-type), 2019~SC9 (V-type), and 2022~UN (C/X-complex) received duplicate classifications with 100\% taxonomic probability in each independent observation. The remaining eight objects (2020~BN12, 2020~LG2, 2020~MX, 2020~NA, 2020~SB, 2020~TP, 2021~BS3, and 2023~DZ2) resulted in different classifications when each observing epoch was analyzed independently. Six of these objects alternated between C/X- or S-complex classifications to A- or D-type classifications, or between D- and A-types. In contrast, 2021~BS3 and 2023~DZ2 were classified as S-type when observed at lower phase angles and brighter apparent magnitudes, and as C/X-complex objects under fainter observing conditions. One possible explanation for these classification changes is that photometric measurements obtained under different observing geometries and signal-to-noise conditions may alter the derived colors and, consequently, the taxonomic classification. For example, observations acquired at brighter conditions could enhance the detectability of olivine and pyroxene absorption features, favoring an S-type classification.}

{Different grain sizes \citep{Bowen_2023} and phase angles \citep{Sanchez2012} are known to cause variability in spectral properties such as slope and absorption features.} We highlight that our training classification sample is based on a taxonomic scheme \citep{Tholen1984, Bus2002, Demeo2009} established from the properties of larger (brighter) bodies in near-Earth space and the main belt. For example, a large asteroid (i.e., D $>10$~km) could hold a highly space weathered regolith composed of smaller grain sizes than a small, younger (fresh surface) and rapidly rotating one. Thus, awareness should be taken about the possibility that the current taxonomy could be inhibiting the proper classifications of certain small NEOs.

\subsection{Biases toward high albedo objects \label{sec:debias}}

The aim of this section is to describe the debiasing method used to correct the taxonomic distribution of our survey with respect to size. We adopted the approach shown by \citet{STUART2004295} and \citet{Marsset_2022} to debias the taxonomic fraction by determining the true intrinsic correction factor between two observed populations (i.e., $Y$ and $Z$). The approach is given by

\begin{equation}
    R_{true} = \frac{N_{true,Y}}{N_{true,Z}} = \frac{N_{obs,Y}}{N_{obs,Z}} \left(\frac{p_{V,Z}}{p_{V,Y}}\right)^{\frac{1}{2}(\alpha - 1)}
\end{equation}
where $R_{true}$ corresponds to the true intrinsic magnitude-limited ratio between the observed number of two populations (i.e., $Y$ and $Z$) with distinct albedos (i.e., $p_{V,Y}$ and $p_{V,Z}$) but similar size and radial distributions. To determine $R_{true}$, we used the power-law slope $\alpha$ of the size frequency distribution (SFD) for objects smaller than 40~m ($\alpha=3.00$), between 40 and 400~m ($\alpha=2.70$), and larger than 400~m ($\alpha=2.65$) from \citet{Heinze_2021}. We used an average $p_{V}$ of 0.04, 0.24, 0.04, 0.26, and 0.34 for C-/X-complex, S-complex, D-types, A-types, and V-types, respectively. We discuss the results of this approach in section \ref{sec:discussion}.

\begin{figure*}[!]
    \centering
    \includegraphics[width=\textwidth]{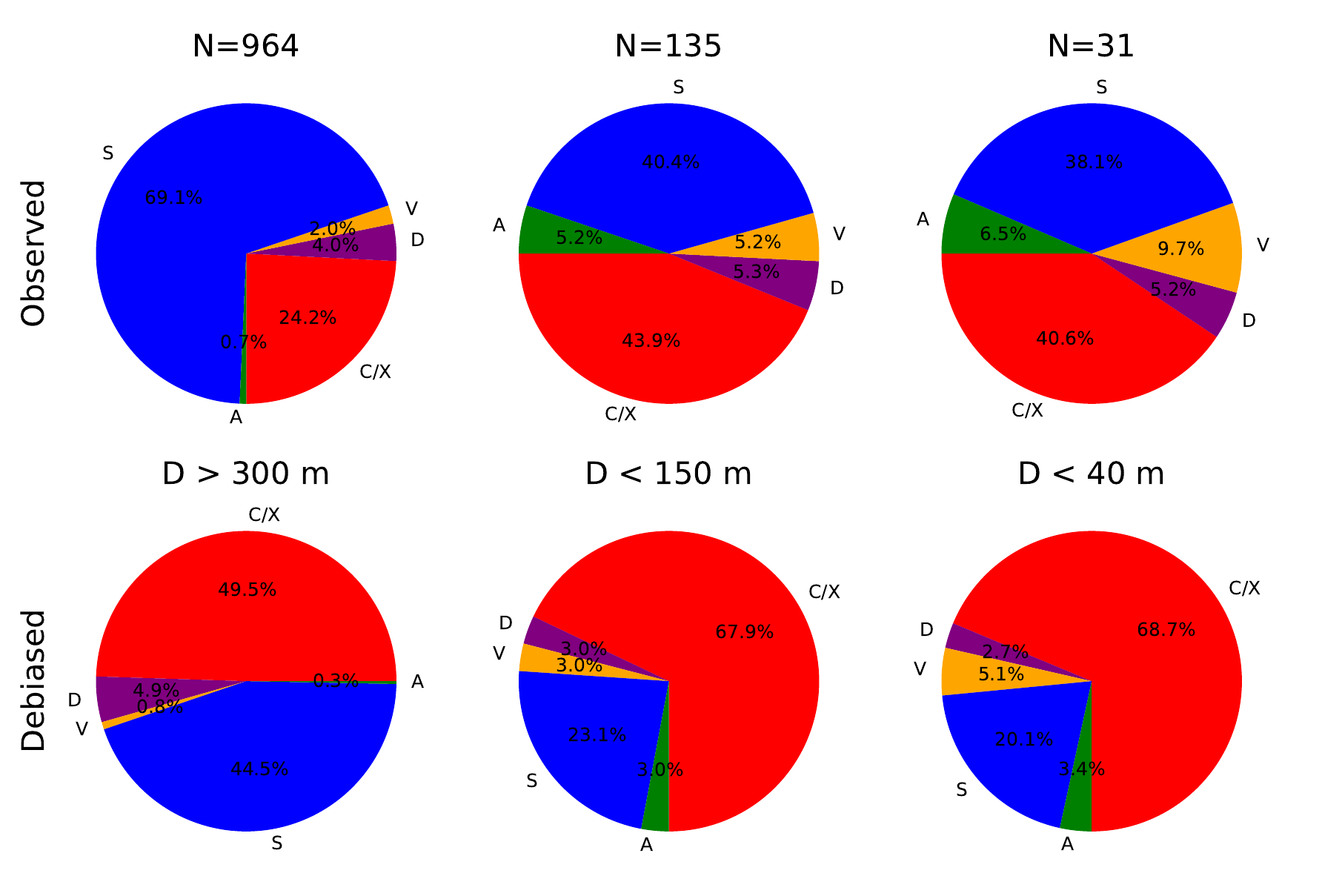}
    \caption{The observed (top) and debiased (bottom) fractional taxonomic distributions are illustrated for three size bins denoted in between the top and bottom pie charts. The letter N in the top indicate the number of asteroids included in each bin. The blue, red, purple, green, and yellow colors represent the fraction of S, C/X, D, A, and V objects per bin. The left pie-chart illustrate the taxonomic fractions of the MITHNEOS survey obtained from \citet{Marsset_2022}. The center- and right-most pie-charts shows the taxonomic fractions obtained in this work.}
    \label{fig:composition_distribution}
\end{figure*}

We stress that this first-order approach relies on a number of simplifying assumptions \citep{Marsset_2022} and ignores additional biases affecting NEO surveys \citep{Devogele2019}. We also assumed that all the objects contained within the C/X-complex sample had surface albedos similar to C-complex (i.e., low albedo). In addition, we assume a similar SFD for all taxonomic types, which is not strictly valid, as younger main-belt families dominate the flux of very small NEOs \citep{Broz2024}. Consequently, the debiased fractions should be interpreted with caution, given the assumptions underlying this technique.

Statistically, NEO surveys preferentially detect objects on low-semimajor-axis orbits, primarily those delivered through the $\nu_{6}$ resonance, because these objects spend a larger fraction of their orbital lifetime close to Earth and are therefore more readily observable than dynamically young NEOs recently injected from the middle and outer main belt, which still retain relatively large semimajor axes. The latter are often detectable only during the brief intervals when they approach perihelion, resulting in significantly lower discovery and characterization probabilities. Consequently, these source populations are likely underrepresented in current NEO surveys. Since many major families in the middle and outer main belt are dominated by low-albedo B-, C-, and X-type asteroids \citep[e.g., Themis, Veritas, Theobalda, Euphrosyne, Ursula, Alauda, Hygiea, Hoffmeister;][]{Broz2024}, this orbital selection effect may further contribute to an underestimation of the abundance of primitive asteroids in the observed NEO population.

From the orbital properties themselves, S-complex objects are more likely to have higher frequencies of being detected from Earth than C-complex objects. Thus, it is expected that the fraction of dark objects will increase even more when also counting the orbital bias. Although we are not including any orbital debias correction in our observed taxonomic distribution, the analytical approach shown in section \ref{sec:append} illustrates another parameter that further favors the detection of S-complex objects based on their orbital parameters and the higher frequency of close encounters with Earth. 

\subsection{The Observed and Debiased Taxonomic Distribution}

The observed and debiased taxonomic distributions as a function of object size are presented in Figure~\ref{fig:composition_distribution}. Our observed distribution is broadly consistent with previous spectroscopic and spectrophotometric surveys \citep[e.g.,][]{Mommert2016, Perna2018, Devogele2019, Moskovitz_2026, Ngwane_2026}, confirming that the population of small NEOs is dominated by silicate-rich S-complex and carbonaceous/metallic C/X-complex asteroids, which together account for the vast majority of the sample and are present in broadly comparable proportions, with the C/X-complex being slightly more abundant (43.9$\pm$7\%). Furthermore, the observed taxonomic fractions remain largely invariant across the size range sampled in this work (see Figure~\ref{fig:composition_distribution}), indicating no strong dependence of taxonomy on object size within the sub-150~m regime.

A comparison with larger NEO surveys, however, reveals a notable difference. The $\sim$70\% fraction of S+Q-type asteroids reported for kilometer-sized NEOs by \citet{Binzel2019} and \citet{Marsset_2022} is substantially higher than the $40.4\pm7$\% measured here for objects with D$<150$~m, suggesting a relative depletion of S-complex bodies toward smaller sizes. A similar trend was reported by \citet{Moskovitz_2026}, who found that the fraction of S+Q objects decreases with decreasing size while the X-complex (analogous to our C/X-complex) becomes increasingly abundant. In contrast, \citet{Sanchez2024} found that the fraction of S-complex asteroids smaller than 300~m is comparable to that of kilometer-sized NEOs. One possible explanation for this discrepancy is that \citet{Sanchez2024} included the recently introduced Sx subclass, which the authors argued represents S-like objects whose relatively weak absorption bands can otherwise lead them to be classified as members of the C- or X-complex. Our measured C/X-complex fraction of $44\pm6$\% is slightly higher than the $\sim$36$\pm$4\% reported by \citet{Moskovitz_2026}, and substantially higher than the 27\% reported by \citet{Perna2018}.

\begin{figure*}[!]
    \centering
    \includegraphics[width=0.8\textwidth]{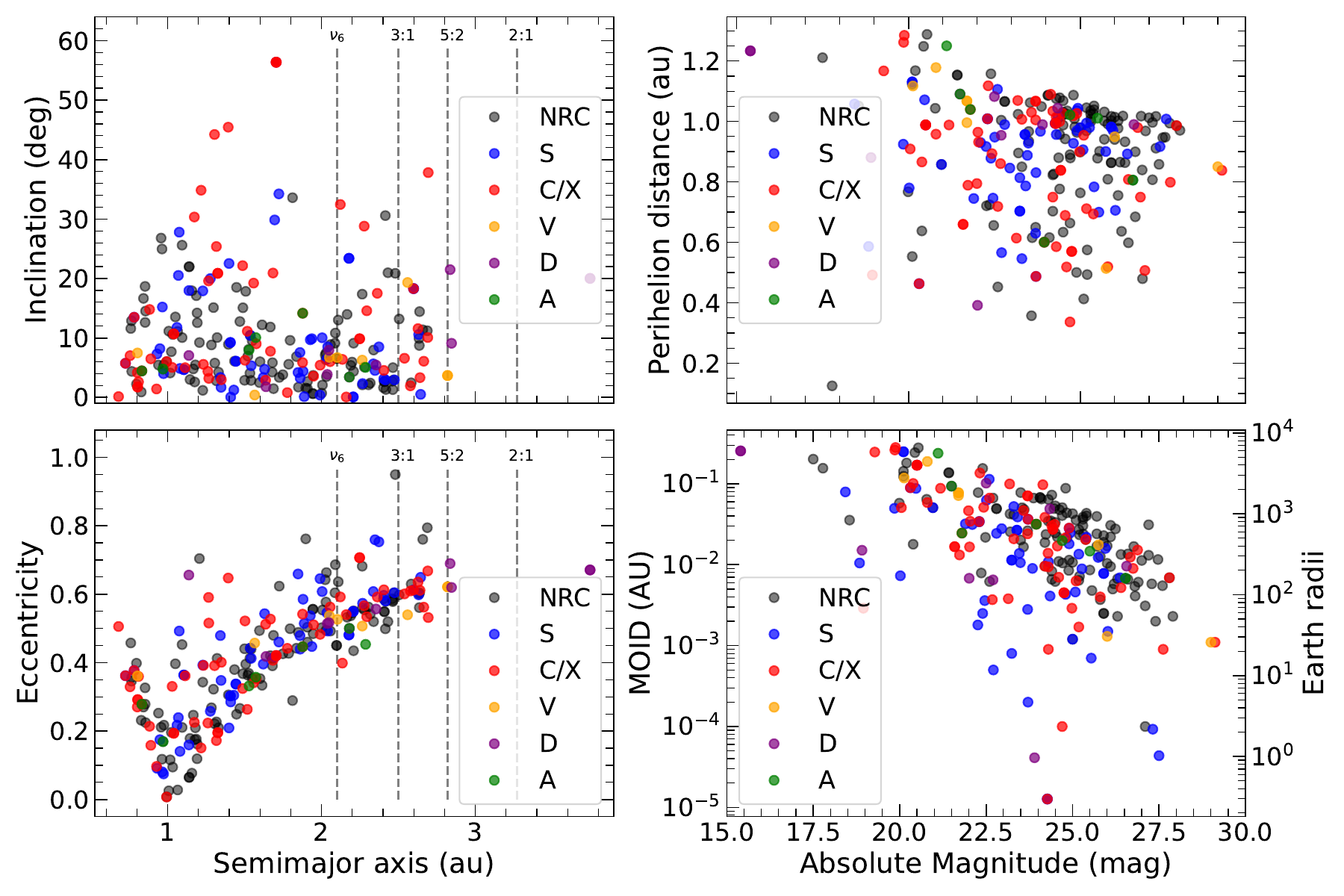}
    \caption{Orbital element distribution for all classified NEOs in our sample. The left panels illustrate the inclination and eccentricity as a function of semi-major axis. The right panels shows the minimum orbit intersection distance (MOID) and perihelion distance as a function of absolute magnitude. The sample spans the canonical near-Earth orbital parameter space, with a concentration at low inclinations. Seven objects have relatively low MOIDs, which if we assume a conservative five Earth radii limit as the minimum distance an object has to fly-by the Earth to experience tidal forces, then they may have experienced regolith resurfacing \citep{Binzel2019}. The objects in black, labeled as NRC, refer to those for which we could not provide reliable classification.}
    \label{fig:orb_param}
\end{figure*}

We observe a possible excess of D-types ($5.3\pm3.0$\%). Interestingly, \citet{Perna2018} found a similar fraction of D-types (7\%) among small NEOs, while \citet{Marsset_2022} also reported a comparable excess ($\sim$5\%) in the MITHNEOS survey, which primarily sampled kilometer-sized objects. These authors proposed the possible existence of an undiscovered family in the middle Main Belt, between the 5:3J and 3:1J mean-motion resonances, that could be responsible for supplying the NEO population with D-type objects. In contrast, the MANOS survey did not find a comparable excess of D-types \citep{Devogele2019, Moskovitz_2026}. We note that our survey covers the near-infrared wavelength range, where D-type asteroids are more readily distinguishable than in the visible. Therefore, it is possible that our survey is sampling a characteristic of the small NEO population that is not as effectively captured by visible-only surveys.

{Additionally, our observed A-type fraction of $5.2\pm3.0$\% is consistent with the value reported by \citet{Perna2018}, although it is nearly a factor of two lower than that measured by \citet{Moskovitz_2026}. In contrast, \citet{Marsset_2022} found a very low A-type fraction ($\leq0.7$\%) among kilometer-sized NEOs, suggesting that these objects may become more abundant toward smaller sizes. The recent discovery of a previously unknown A-type family in the Main Belt \citep{Galinier2024}, together with revised estimates of the intrinsic abundance of A-type asteroids \citep{Delbo2026}, indicates that these objects are more common than previously believed. Therefore, the A-type abundance measured in our survey, which is nearly a factor of two higher than the latest estimates from Gaia \citep{Delbo2026}, may reflect an intrinsic property of the very small NEO population. If confirmed with larger samples, these results would imply that both D- and A-type asteroids are more prevalent among the smallest NEOs than previously recognized, providing new constraints on source-region models and the compositional evolution of material delivered from the Main Belt (see Section~\ref{sec:source}).}

After debiasing the observed taxonomic distribution, the overall fraction of S-complex bodies decreases from 40\% to 23\%, while the C/X-complex fraction increases from 44\% to 67\%. The other taxonomic types explored here (i.e., A-, D-, and V-types) show relatively equal abundances of $\sim$3\% each. As mentioned in Section~\ref{sec:debias}, the debiasing technique is a first-order approach that relies on several assumptions, and additional factors beyond albedo bias must be taken into account. However, the fairly good agreement between the ratio of S-complex to V-type objects from the debiased distribution ($\sim$11) and the ratio of ordinary chondrites (OC) to Howardite–Eucrite–Diogenite (HED) meteorites ($\sim$10–11) derived from meteorite fall statistics \citep{Gerome2025} serves as a possible validation, assuming that these objects are equally filtered during atmospheric entry. 

If the debiased distributions shown in Figure~\ref{fig:composition_distribution} more closely reflect the true composition of the small NEO population, several important implications follow. First, they imply that small NEOs are predominantly primitive (dark) rather than silicate-rich (bright). Recent work by \citet{Broz2024} showed that a large fraction of meteorites originates from young ($<$8~Myr) collisional breakups in the Koronis and Massalia families. Although these families are primarily associated with S-complex asteroids, the authors also concluded that the CM-like Veritas family \citep{Farley2006, Vernazza2016}, formed approximately 8.3~Myr ago \citep{Nesvorny2003}, should be contributing about the same order as the total meteorite flux \citep{Nesvorny2023, Brovz2024b}. This further supports a scenario in which the NEO population is dominated by primordial objects that remain largely undetected due to the detection limits of current telescopic facilities. Independent support for this interpretation comes from \citet{Shober2025}, whose dynamical models of the near-Earth meteoroid population predict that more than half of the incoming material should share origins with carbonaceous chondrites based on asteroid belt population estimates. However, it remain uncertain whether these models of meteoroid could be used to extrapolate properties of the NEO population between 20 to 150~m.

Finally, meteorite fall statistics indicate that ordinary chondrites outnumber carbonaceous chondrites ($\sim$4.4\%) by about a factor of 18, and by about a factor of 53 when only CM meteorites ($\sim$1.5\%) are considered. Comparing these values with our debiased C/X-complex fractions suggests that carbonaceous material entering Earth’s atmosphere is reduced by at least a factor of 15 relative to the number of recovered samples. This result, albeit lower, is consistent with recent meteoroid and meteorite modeling studies \citep{Broz2024, Shober2025}. Due to the relatively low tensile strength of carbonaceous materials, a large fraction of the incoming mass can be destroyed at very low dynamic pressures in the upper atmosphere, resulting in only a small fraction surviving to reach the ground \citep{Popova2005, Borovicka2019}. In addition, \citet{Shober2025} found that thermal processing \citep{Delbo2014} in space prior to atmospheric entry—particularly at low perihelion distances— weakens fragile meteoroids, causing them to fragment and break apart, thereby reducing their likelihood of reaching Earth’s atmosphere.

\subsection{Dynamical origins and link to meteorites \label{sec:source}}

Tracing NEOs back to their source regions in the main belt and linking them to specific meteorite types is essential for understanding the delivery pathways of meteoritic material to Earth and for placing laboratory meteorite analyses within a broader dynamical and compositional context. Such connections provide a physically motivated framework for interpreting NEO taxonomic surveys and for assessing the representativeness of meteorite collections relative to their parent populations.

In Figure~\ref{fig:orb_param}, we present the orbital elements of the NEOs with robust probabilistic taxonomic classifications. Rather than identifying the escape resonance through which each NEO was delivered to near-Earth space \citep{Granvik2018, Nesvorny2023}, we estimated the most probable source family of each object using the recently developed METEOMOD orbital distribution model \citep{Broz2024}, which accounts for contributions from 52 asteroid families. Because METEOMOD was developed to reproduce the source distribution of meter-sized meteoroids, substantially smaller than the $\sim$30--130~m objects considered in this work, its source probabilities may be biased toward younger families whose size-frequency distributions dominate the meteoroid flux. To assess the robustness of the inferred source regions, we also applied the NEOMOD model \citep{Nesvorny2023} following the same methodology as \citet{Broz2024}. NEOMOD was developed to better reproduce the orbital distribution of $\sim$100~m NEOs and therefore provides a more appropriate comparison for our sample. We note, however, that the size-frequency distributions of most candidate source families remain poorly constrained at diameters of tens to hundreds of meters. Consequently, there is still much to be explored regarding the dynamical pathways and family-level origins of small NEOs. For the purposes of this analysis, we assigned each object to the family with the highest source probability, irrespective of the absolute probability, recognizing that only $\sim$20\% of the sample yielded maximum source probabilities exceeding 50\%.

{The METEOMOD model suggests that the majority of objects in our sample (54\%) originate from the Massalia family \citep{Marsset2024}, followed by Koronis (22\%), Veritas (10\%), Phaethon (4\%), König (3\%), Hungaria (2\%), and Polana (2\%), with Hoffmeister, Flora, Aeolia, Adeona, Vesta, and Misa accounting for the remaining fraction. This result is broadly consistent with the expected dominance of ordinary-chondrite-producing families in the meteoroid flux \citep{Marsset2024, Broz2024}. In contrast, the NEOMOD model indicates that nearly 80\% of the objects in our survey have the highest dynamical affinity with the Flora family, followed by Polana (15\%), Adeona (2.5\%), and Koronis, Hoffmeister, Phocaea, Vesta, and Phaethon, each contributing less than 1\%. The predominance of Flora is expected given its proximity to the $\nu_{6}$ secular resonance, one of the most efficient dynamical pathways for delivering inner Main Belt asteroids into near-Earth space \citep{Vernazza2008, Marsset_2022}.}

{The markedly different source distributions inferred by METEOMOD and NEOMOD emphasize that the dynamical origins of $\sim$100~m NEOs remain incompletely understood and likely depend on the size regime for which the models were developed. Nevertheless, both models preferentially associate our sample with families traditionally linked to S-complex asteroids, whereas our observed taxonomic distribution indicates nearly equal proportions of S- and C/X-complex objects. One possible explanation for this apparent discrepancy is that small asteroids experience enhanced Yarkovsky-driven mobility, allowing them to drift across a broader region of the inner Main Belt before entering the $\nu_{6}$ and 3:1 resonances \citep{Bottke2006, 2017A&A...598A..52G}. Such increased mobility would naturally sample a wider range of taxonomic environments and could produce the greater compositional diversity observed among the smallest NEOs, as previously suggested by \citet{Moskovitz_2026}. We also note that the enhanced abundance of primitive (C/X-complex) objects among the smallest NEOs could additionally reflect evolutionary processes such as thermal-induced fragmentation \citep{LATSIA2026117130} and tidal disruption during close encounters with terrestrial planets \citep{Granvik_2024}, both of which may preferentially increase the number of small bodies within the NEO population. Although investigating these processes is beyond the scope of this work, the observed taxonomic distributions presented here provide empirical constraints that future dynamical models may use to refine and validate the delivery mechanisms of small NEOs from the Main Belt.}

{We note that for one object in our sample, 2021~PE20, we could not identify a likely main-belt source region using the METEOMOD or NEOMOD models. 2021~PE20 was independently classified as a D-type in two separate observing epochs, each with a taxonomic probability of $\geq$50\%. Using the NEO source model of \citet{Granvik2018}, we estimate a 97\% probability that 2021~PE20 originated from the Jupiter Family Comets (JFCs). A JFC link is further supported by reports of cometary activity in the days following its discovery, which cataloged the object as P/2021 PE20. In addition, our $J$-band thumbnail images obtained on 2021 August 30 and September 3 appear to show possible evidence of activity.}

\section{Conclusion}

We conducted 230 rapid-response near-infrared spectrophotometric taxonomic characterization of recently discovered small NEOs. Our campaign specifically targeted objects smaller than 130~m in diameter, yielding a mean diameter of approximately 60~m. By sampling this size regime, we directly probed the population most likely to represent the precursors of meteors and meteorites reaching Earth. The observed taxonomic fractions are broadly consistent with those reported in previous taxonomic studies, suggesting that silicate-rich and carbonaceous/metallic like asteroids are represented at almost equal proportions in the population of small NEOs. In addition, our results provide further evidence for a size dependence in the taxonomic distribution of NEOs, suggesting that distinct processes may influence the relative abundances of the smallest objects. A plausible explanation for the observed abundances of primitive NEOs is the enhanced Yarkovsky-driven mobility of small asteroids within the Main Belt in combination with thermal-induced fragmentation and tidal disruption during close encounters with terrestrial planets. Our debiased taxonomic distributions—supported by meteorite fall statistics and meteoroid population models—suggest that the intrinsic NEO population should be even more dominated by primordial,  C/X-complex objects. Furthermore, our results imply that a substantial fraction of fragile carbonaceous meteoroids does not survive atmospheric entry. This finding underscores the critical importance of sample-return missions for advancing our understanding of these primitive extraterrestrial materials. 

\textit{Recommendations for future taxonomic surveys.} The results and methodologies presented in this work underscore the critical importance of debiasing taxonomic surveys. Although challenging, addressing observational and selection biases is essential for achieving a realistic understanding of the relationships among NEOs, meteors, and meteorites. A key advancement for future taxonomic characterization—particularly for small, fast-rotating, and irregularly shaped asteroids— will be the acquisition of visible to near-infrared spectroscopy to enable the distinction of taxonomies by spectral features that are not accessible via spectrophotometry. Additionally, surveys with pointing history and well characterized detection efficiencies to incorporate in survey simulators is critical.

\begin{longrotatetable}
\begin{deluxetable*}{ccccccccccccccccc}
\tablecaption{Results of the UKIRT NEO survey. \label{tab:results}}
\tablewidth{0pt}
\tablehead{
\colhead{Object} & \colhead{Date} & \colhead{$H_{V}$} & \colhead{$V$} & \colhead{$J_{median}$} & \colhead{$Z-J$} & \colhead{$J-H$} & \colhead{$J-K$} & \colhead{$A^{*}$} & \colhead{$\tau^{*}$} & \colhead{C/X} & \colhead{S} & \colhead{V} & \colhead{D} & \colhead{A} & \colhead{$Tax.$} & \colhead{Source} \\
\colhead{} & \colhead{(UT)} & \colhead{(mag)} & \colhead{(mag)} & \colhead{(mag)} & \colhead{(mag)} & \colhead{(mag)} & \colhead{(mag)} & \colhead{(mag)} & \colhead{(minutes)} & \colhead{Prob.} & \colhead{Prob.} & \colhead{Prob.} & \colhead{Prob.} & \colhead{Prob.} & \colhead{} & \colhead{} \\
}
\startdata
2016 YS & 16-12-23 10:46 & 25.0 & 16.64 & 14.98 & 0.08$\pm$0.1 & 0.12$\pm$0.07 & -0.04$\pm$0.07 & 0.49 & 18.7 & 0.21 & \textbf{0.79} & 0.0 & 0.0 & 0.0 & S & Polana (CI) \\
2016 WW9 & 16-12-28 14:37 & 18.91 & 18.66 & 17.75 & 0.27$\pm$0.1 & 0.21$\pm$0.07 & 0.56$\pm$0.11 & 0.44 & 39.6 & 0.0 & 0.01 & 0.0 & \textbf{0.79} & 0.2 & D & Brasilia (M) \\
2017 AG5 & 17-01-09 08:47 & 22.4 & 16.61 & 15.68 & 0.04$\pm$0.02 & 0.22$\pm$0.03 & 0.19$\pm$0.03 & 0.03 & 64.8 & 0.16 & \textbf{0.83} & 0.0 & 0.0 & 0.0 & S & Flora (LL) \\
2017 BW & 17-01-29 08:26 & 23.49 & 18.48 & 17.45 & 0.06$\pm$0.07 & 0.24$\pm$0.09 & 0.37$\pm$0.16 & 0.18 & 6.8 & 0.13 & 0.43 & 0.0 & 0.29 & 0.15 & S & Polana (CI) \\
2016 XH1 & 16-12-15 09:28 & 20.39 & 18.32 & 17.12 & 0.04$\pm$0.04 & 0.09$\pm$0.06 & 0.26$\pm$0.06 & 0.06 & 65.7 & \textbf{0.84} & 0.04 & 0.0 & 0.13 & 0.0 & CX & Flora (LL) \\
2017 MB1 & 17-06-27 11:47 & 18.84 & 18.45 & 16.92 & 0.71$\pm$0.06 & -0.1$\pm$0.05 & 0.13$\pm$0.1 & 0.11 & 32.1 & 0.0 & \textbf{0.55} & 0.44 & 0.0 & 0.0 & S & Polana (CI) \\
2017 LD & 17-06-05 12:57 & 27.5 & 18.5 & 16.84 & 0.16$\pm$0.12 & 0.37$\pm$0.1 & 0.16$\pm$0.11 & 0.45 & 6.8 & 0.02 & \textbf{0.94} & 0.0 & 0.04 & 0.01 & S & Flora (LL) \\
2017 MC4 & 17-06-29 11:52 & 22.01 & 19.05 & 17.61 & 0.18$\pm$0.06 & 0.15$\pm$0.06 & 0.59$\pm$0.09 & 0.6 & 14.3 & 0.0 & 0.0 & 0.0 & \textbf{0.86} & 0.14 & D & Flora (LL) \\
2017 MY2 & 17-06-28 13:31 & 21.11 & 20.33 & 18.13 & 0.11$\pm$0.06 & 0.29$\pm$0.07 & 0.52$\pm$0.12 & 0.28 & 75.6 & 0.01 & 0.18 & 0.0 & 0.28 & \textbf{0.53} & A & Flora (LL) \\
2017 JV2 & 17-05-21 09:08 & 22.59 & 18.48 & 17.04 & 0.04$\pm$0.07 & 0.16$\pm$0.06 & 0.36$\pm$0.14 & 0.17 & 21.0 & 0.35 & 0.38 & 0.0 & 0.24 & 0.03 & S & Polana (CI) \\
... & ... & ... & ... & ... & ... & ... & ... & ... & ... & ... & ... & ... & ... & ... & ... & ... \\
2022 SJ9 & 22-09-28 08:37 & 27.32 & 18.64 & 16.91 & 0.17$\pm$0.06 & 0.4$\pm$0.07 & 0.23$\pm$0.08 & 0.37 & 21.1 & 0.0 & \textbf{0.98} & 0.0 & 0.02 & 0.0 & S & Polana (CI) \\
2023 PZ & 23-08-15 10:48 & 29.12 & 18.39 & 17.07 & -0.34$\pm$0.07 & 0.03$\pm$0.09 & 0.41$\pm$0.09 & 0.75 & 28.1 & \textbf{0.96} & 0.04 & 0.0 & 0.0 & 0.0 & CX & Flora (LL) \\
2023 MD2 & 23-07-12 13:34 & 24.35 & 17.43 & 15.84 & 0.27$\pm$0.05 & 0.12$\pm$0.03 & 0.15$\pm$0.1 & 0.45 & 13.5 & 0.02 & \textbf{0.6} & 0.0 & 0.35 & 0.03 & S & Polana (CI) \\
2022 RP2 & 22-09-09 11:30 & 22.02 & 18.51 & 17.14 & 0.16$\pm$0.05 & -0.07$\pm$0.06 & 0.22$\pm$0.06 & 0.15 & 21.1 & \textbf{0.85} & 0.01 & 0.0 & 0.1 & 0.04 & CX & Phocaea (H) \\
2023 DZ2 & 23-03-25 08:50 & 24.27 & 12.6 & 10.04 & 0.74$\pm$0.06 & 0.06$\pm$0.06 & 0.3$\pm$0.09 & 0.56 & 42.3 & 0.0 & \textbf{0.81} & 0.01 & 0.18 & 0.0 & S & Flora (LL) \\
2023 OQ3 & 23-07-29 11:17 & 25.08 & 17.22 & 15.8 & 0.1$\pm$0.03 & 0.18$\pm$0.04 & 0.25$\pm$0.03 & 0.1 & 21.4 & 0.29 & 0.46 & 0.0 & 0.25 & 0.0 & S & Flora (LL) \\
2023 XJ2 & 23-12-08 07:49 & 24.3 & 16.08 & 14.88 & -0.15$\pm$0.02 & 0.12$\pm$0.03 & 0.18$\pm$0.02 & 0.16 & 18.1 & \textbf{1.0} & 0.0 & 0.0 & 0.0 & 0.0 & CX & Flora (LL) \\
\enddata
\tablecomments{The full version of this table is available in machine-readable format in the online journal. A portion is shown here for guidance regarding its form and function. Bolded values indicate probabilistic taxonomic solutions with probabilities greater than 0.5. The letters in parentheses under the source column indicate the associated meteorite affinity of each family.}
\end{deluxetable*}
\end{longrotatetable}

\acknowledgements 
{The authors would like to express their sincere gratitude to an anonymous reviewer for their exceptionally thoughtful and insightful comments, which substantially strengthened both the presentation and the scientific interpretation of this work.} This material was in part supported by the National
Science Foundation Graduate Research Fellowship Program under grant No.\ 2021318193 to ALO. Any opinions, findings, conclusions, or recommendations expressed in this material are those of the author(s) and do not necessarily reflect the views of the National Science Foundation. ALO was in part supported by an appointment to the NASA Postdoctoral Program at the NASA Goddard Space Flight Center, administered by Oak Ridge Associated Universities under contract with NASA.
Part of the computational analyses was carried out on Northern Arizona University's Monsoon computing cluster, funded by Arizona's Technology and Research Initiative Fund.

\appendix

\section{Orbital debiasing \label{sec:append}}

We developed an analytical framework to quantify the fraction of time that an object remains above the survey detection threshold ($V\leq V_{\rm lim}=22$~mag) relative to periods when it falls below detectability over an arbitrary orbital interval. The observability of a set of selected objects with 100\% probabilistic taxonomy solution (i.e., 2016~WW9, 2018~ED1, 2019~UX9, 2020~OJ2, 2020~KR1, and 2020~YQ3) was modeled using an $N$-body orbital integration performed with the REBOUND package \citep{Rein2012}. The dynamical model included the Sun and all major planets, allowing the osculating elements to be propagated forward in time and the observing geometry to be computed as seen from UKIRT. Each NEO orbit was initialized from its Keplerian elements ($a$, $e$, $i$, $\Omega_{0}$, $\omega$, and $M_{0}$).

We considered an object $\mu$ with geometric albedo $p_V$ and adopted the $H$–$G$ photometric system \citep{Bowell1989} to compute the apparent magnitude $V_\mu$,

\begin{equation}
V_{\mu}=C+5\log_{10}(r\Delta)-2.5\log_{10}(D^{2}p_V)-2.5\log_{10}(\Theta),
\end{equation}
evaluated at 1~hr cadence over a 10$^{4}$~yr interval. Here $\Theta=(1-G)\Phi_1(\alpha)+G\Phi_2(\alpha)$ is the phase function, where $G$ is the phase-slope parameter and $C$ is a constant defining the absolute-magnitude normalization \citep{STUART2004295}.

We defined a Heaviside step function $U$ to construct a time-indicator function,
\begin{equation}
\tau_\mu(t)=U(V_{\rm lim}-V_\mu)\cdot U(\zeta(t)),
\end{equation}
which equals unity when the brightness and observing constraints $\zeta(t)$ (see Section~\ref{sec:observation}) are satisfied. The observable time fraction was then computed as

\begin{equation}
T_\mu=\frac{1}{t_1-t_0}\int_{t_0}^{t_1}\tau_\mu(t),dt.
\end{equation}

To obtain a symmetric measure of relative observability, we define, 
\begin{equation}
\varphi=\ln\left(\frac{T_C}{T_S}\right),
\end{equation}
using representative C- and S-complex objects with albedos appropriate to their taxonomic classes. The resulting simulations indicate $\varphi\sim43$, implying that C-complex objects spend roughly forty times longer below the survey detection limit than S-complex objects due to the combined effects of orbital geometry and surface reflectance. In logarithmic space, C-complex objects show a relative excess of $\sim$3.76 toward non-observable conditions compared to S-complex objects, indicating a substantially higher fraction of time spent below the detection threshold.

We emphasize that this metric is inherently orbit dependent, as the fraction of time spent above or below the detection threshold varies strongly with orbital geometry, phase angle evolution, and viewing circumstances. This approach is not intended to provide a full population debiasing but rather to illustrate the magnitude and direction of observability biases affecting taxonomic fractions. The derived values should therefore be interpreted qualitatively, demonstrating that low-albedo objects can remain undetectable for substantially longer fractions of their orbital evolution, rather than as a quantitative correction applied to the survey results. 

\setcounter{figure}{0}
\renewcommand{\thefigure}{A\arabic{figure}}

\begin{figure}[t!]
    \centering
    \includegraphics[width=0.5\textwidth]{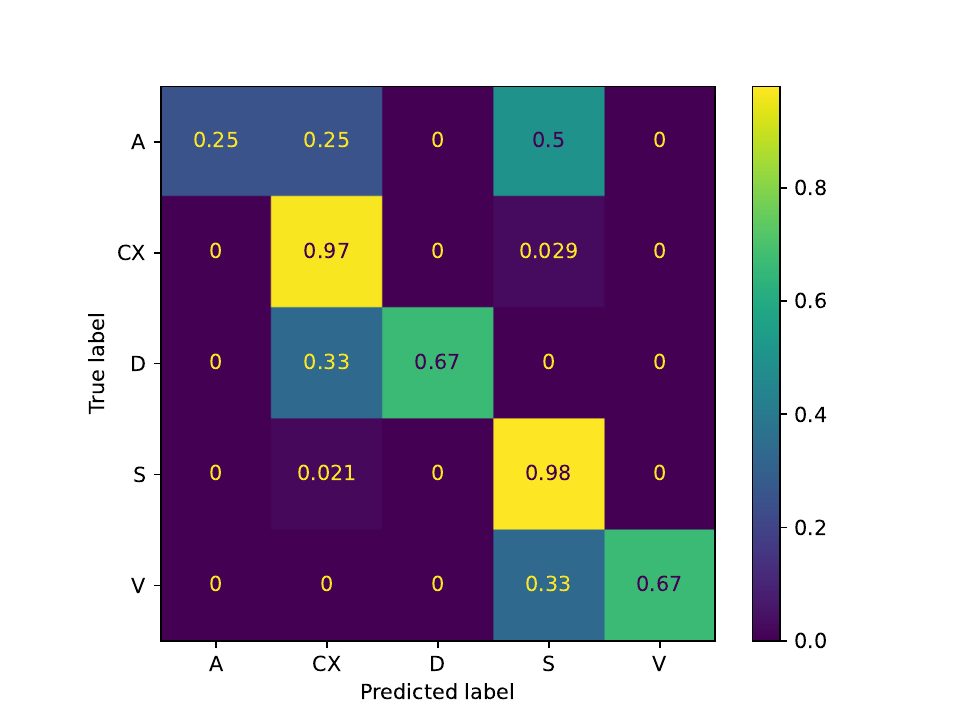}
    \caption{Confusion matrix for three-color classification scheme. The diagonal shows the fraction of objects correctly classified by the training sample shown in Figure \ref{fig:training}.}
    \label{fig:confusion_matric}
\end{figure}

\clearpage
\bibliography{references}{}
\bibliographystyle{aasjournal}

\end{document}